# Considerations about the randomness of bit strings originated from sequences of integer numbers according to a simple quantum computer model

R. V. Ramos

rubens.viana@pq.cnpq.br

*Lab. of Quantum Information Technology, Department of Teleinformatic Engineering – Federal University of Ceara - DETI/UFC, C.P. 6007 – Campus do Pici - 60455-970 Fortaleza-Ce, Brazil.*

This work uses a simple quantum computer model to discuss the randomness of bit strings originated from integer sequences. The considered quantum computer model has three elements: a processing unit responsible for a mathematical operation, an initial equally weighted superposition and a quantum state used as resource. The randomness depends on the last.

## 1. Introduction

Firstly, consider that $S$ is a finite sequence of integer numbers: $S = \{s_1, s_2, \ldots, s_n\}$. Now, let us assume there is a question $Q$ about each element of $S$ whose answer is true or false. Thus, one can build the binary string $B = Q(s_1)Q(s_2)\ldots Q(s_n)$, where $Q(s_i)$ is equal to '1'('0') if $s_i$ (does not) satisfies the question $Q$. The subset of $S$ formed by the elements of $S$ that satisfy $Q$ is the set $S_Q$. The randomness of the binary string $B$ depends on the question $Q$. Two examples of questions are: 1) which elements of $S$ are even? 2) Which elements of $S$ are primes? Chaitin linked the randomness of a bit string to the length of the shortest computer program able to generate it [1,2]. A given bit string is said to have high randomness if it has (roughly) the same length of the shortest computer program able to generate it. In this case, such bit string is said to be incompressible. This idea is explored in this work by using a simple quantum computer model.

Given the (not normalized) quantum state $|S\rangle = \sum_{i=1}^{n}|s_i\rangle$ (related to the set of integer numbers $S = \{s_1, s_2, \ldots, s_n\}$) and the question $Q$, the quantum computer composed by a unitary operation $U_Q$ and the quantum state $|W\rangle = \sum_{i=1}^{m}|w_i\rangle$ (related to the set $W = \{w_1, w_2, \ldots, w_m\}$), is used to mark the elements of $|S\rangle$ that satisfy $Q$: $U_Q|s_i\rangle|w_j\rangle|0\rangle = |s_i\rangle|w_j\rangle|Q(s_i,w_j)\rangle$. Thus, $W$ is the set of integer numbers, with the minimum cardinality, required to answer the question $Q$ for all elements of $S$. If $U_Q|s_i\rangle|w_j\rangle|0\rangle = |s_i\rangle|w_j\rangle|1\rangle$ then $w_j$ is named a witness of $s_i$. The quantum algorithm that marks all the elements of $S$ that satisfy $Q$ is

$$|B\rangle = U_Q \frac{1}{\sqrt{2^{n+m}}}\left[|s_1\rangle+|s_2\rangle+\ldots+|s_n\rangle\right]\left[|w_1\rangle+|w_2\rangle+\ldots+|w_m\rangle\right]|0\rangle = \frac{1}{\sqrt{2^{n+m}}}\sum_{i=1}^{n}\sum_{j=1}^{m}|s_i\rangle|w_j\rangle|Q(s_i,w_j)\rangle.$$
(1)

One may note that having the quantum state (1), quantum search [3] can be used to produce a quantum state whose elements are only the elements of $S_Q$, as well a quantum counting algorithm [4] can be used to determine the cardinality of $S_Q$.

Regarding the randomness defined by Chaitin, the quantum state $|W\rangle$ plays the role of the classical computer program. The quantum state $|W\rangle$ has the required information to generate $|B\rangle$ using $U_Q$ (the quantum state $|S\rangle$ is the input of the quantum computer in the same way that $S$ is the input of the algorithm $C$ in the classical computer). If $|W| = m < q = |S_Q|$, where $|x|$ means the cardinality of the set $x$, then the binary sequence $B$ obtained from $S$ and $Q$ is compressible. The larger the difference $q$-$m$ the larger is the compression. On the other hand, if $q = m$ (both sets have the same cardinality) then the binary sequence $B$ is incompressible and it has the maximum randomness.

As can be observed, the bit sequence $B$ can be compressed if there is at least one element of $W$ that is a witness of at least two elements of $S_Q$. For example, $U_Q|s_i\rangle|w_j\rangle|0\rangle = |s_i\rangle|w_j\rangle|1\rangle$ and $U_Q|s_k\rangle|w_j\rangle|0\rangle = |s_k\rangle|w_j\rangle|1\rangle$. Hence, $w_j$ is witness of $s_i$ and $s_k$. On the other hand, if each element of $S_Q$ has a unique witness, then the number of witnesses is equal to the cardinality of $S_Q$ and the bit sequence $B$ is incompressible. At last, in order to have the set $W$ with the smallest cardinality, an element of $S$ cannot have two (or more) different witnesses. For example, if $U_Q|s_i\rangle|w_j\rangle|0\rangle = |s_i\rangle|w_j\rangle|1\rangle$ and $U_Q|s_i\rangle|w_k\rangle|0\rangle = |s_i\rangle|w_k\rangle|1\rangle$ then $w_j$ and $w_k$ are both witnesses of $s_j$. In this case, one of them can be discarded and the cardinality of $W$ is decreased by one unity.

## 2. Examples

Let us start by considering $S$ the integer numbers from 1 to $n$. The question $Q$ is: which elements of $S$ belong to the sequence $x_{k+1} = px_k+q$, with $x_0=1$? In this case $U_Q$ implements the operation $U_Q|s\rangle|w\rangle|0\rangle = |s\rangle|w\rangle|((s \bmod p)\oplus w)\oplus 1\rangle$ (for simplification it is understood that $(s \bmod p)\oplus w = 0$ if $(s \bmod p) = w$) and the set $W$ is only $W = \{q\}$, what means there is no randomness and the compression is maximal.

Now, we consider the randomness in the prime distribution. The set $S$ is the set of integer numbers from 2 to $n$ (where $n$ is an arbitrarily large number), and the question is: Which elements of $S$ are not prime numbers? The elements of $W$ will be witnesses of composite numbers. Thus, the elements of the set $W$ are the prime numbers from 2 up to $n^{1/2}$. The quantum operation is $U_Q|s\rangle|w\rangle|0\rangle = |s\rangle|w\rangle|(\text{isinteger}(s/w)\rangle$, where the function isinteger($x$) returns '1' ('0') if $x$ is (not) an integer number. Since the number of primes between 2 and square root of $n$ is smaller than the number of composite numbers between 1 and $n$, there is some compression. However, the minimal length of $W$ is $\pi(n^{1/2})$ where $\pi$ is the prime count function, thus, as expected, there is some randomness in the prime distribution.

A more complicate situation occurs when one considers the Möbius function: $\mu(k) = 0$ if $k$ has at least one repeated prime factor, $\mu(1) = 1$ and $\mu(k) = (-1)^l$ when $k$ is a product of $l$ distinct primes. The Möbius function is very important in number theory, for example, it is related to the Riemann hypothesis. Let $S = \{s_1,..,s_n\}$ be the set of the first $n$ (again we consider $n$ an arbitrary large number) integer numbers with Möbius function different of zero, $\mu(s_i) = \pm 1$ for $i \in [1,n]$. The question $Q$ is: Which elements of $S$ have Möbius function equal to $+1$? The bit sequence originate from $S$ and $Q$ is $B = \{0.5[1+\mu(s_1)], 0.5[1+\mu(s_2)],\ldots, 0.5[1+\mu(s_n)]\}$ In this case, the witnesses (the set $W$) are the elements of $S$ with Möbius function equal to -1 and $U_Q|s\rangle|w\rangle|0\rangle = |s\rangle|w\rangle|\text{isprime}(s/w)\rangle$. The function isprime($k$) returns the value 1 (0) if $k$ is (not) a prime number [5]. In fact, for any integer $k$ with $\mu(k) = 1$ there exist an integer $t$ with $\mu(t) = -1$ such that $k/t$ is prime. For example: let $p_1,\ldots,p_{l-1},p_l$ be $l$ distinct prime numbers, where $l$ is an even number. Then $t = p_1 p_2 \ldots p_{l-1}$ ($\mu(t) = -1$) is a witness that $\mu(k) = 1$ for $k = p_1 p_2 \ldots p_l$, since $k/t = p_l$. Furthermore, since $p_l$ can be any prime number (distinct of the first $l$-1 prime numbers considered), $t$ is a witness of many numbers with Möbius function equal to 1. This fact points to the possibility of compression. However, the number $k = p_1 p_2 \ldots p_l$ has other witnesses. For example, $r = p_2 \ldots p_l$ is also a witness of $k$ since $k/r = p_1$. In fact, there are $l$ distinct witnesses for $k$ and none of them can be discarded, since each one is witness of other elements. For example, the number 35 has two witnesses: 5 and 7. Let us discard the number 5. The number 21 has also two witnesses: 3 and 7. Since the number 7 cannot be discard (otherwise 35 would not have any witness), the number 3 is discarded. However, now the number 15 has not witnesses, since 3 and 5 were discarded. Therefore, there is a kind of paradox since any element of $W$ is witness of several element of $S_Q$, what would mean the possibility of compression ($|W|<|S_Q|$), and any element of $S_Q$ has more than one witness, pointing for an excess of witnesses, ($|W|>|S_Q|$). In this case, the randomness cannot be checked. By inspection, one may note that using as witnesses the elements of $S$ with Möbius function equal to $+1$ the same problem would appear. A solution of this paradox is to make $W = S_Q$ (and, hence, $|W| = |S_Q|$), that is, each element of $S$ that satisfies $Q$ is a witness of itself. The quantum operation is simply $U_Q|s\rangle|w\rangle|0\rangle = |s\rangle|w\rangle|(s\oplus w)\oplus 1\rangle$. Since $|W| = |S_Q|$, the bit sequence $B = \{0.5[1+\mu(s_1)], 0.5[1+\mu(s_2)],\ldots, 0.5[1+\mu(s_n)]\}$ has maximal randomness and it is not compressible.

## 3. Conclusions

This work tries to answer the following question: given the (not normalized) states $|S\rangle = \sum_{i=1}^{n}|s_i\rangle$ and $|S_Q\rangle = \sum_{i=1}^{l}|s_{Qi}\rangle$, where each $S_{Qi}$ is an element of $S$ that satisfies a given question $Q$, how much information is required by the quantum circuit that realizes the operation $U|S\rangle = |S_Q\rangle$? In order to get this answer, we divided $U$ in two parts: a unitary operation $U_Q$ and a (not normalized) quantum state $|W\rangle = \sum_{i=1}^{m}|w_i\rangle$. The last carries the amount of information required by $U$. Since $|W\rangle$ is linked to the question $Q$ and, hence, to $|S_Q\rangle$, the randomness, that depends on $Q$, depends also on the relation between the cardinalities of $W$ and $S_Q$ or, equivalently, depends on the relation between the number of terms of $|W\rangle$ and $|S_Q\rangle$.

As written before, using quantum search to select only the marked states ($Q(s_i,w_j) = 1$) in (1) one can get the quantum state whose elements are only the elements of $S_Q$, $|S_Q\rangle$. Now, let us assume that $|S_Q| = l$. Considering the randomness one has three possibilities for the state $|S_Q\rangle$:

$$I)\,|S_Q\rangle = \frac{1}{\sqrt{2^l}}\left[|s_{Q1}\rangle + |s_{Q2}\rangle + \ldots + |s_{Ql}\rangle\right]|w\rangle \qquad (2)$$

$$II)\,|S_Q\rangle = \frac{\left[|s_{Q1}\rangle + \ldots + |s_{Qk}\rangle\right]}{\sqrt{2^k}}|w_1\rangle + \frac{\left[|s_{Qk+1}\rangle + \ldots + |s_{Ql}\rangle\right]}{\sqrt{2^{l-k}}}|w_2\rangle \qquad (3)$$

$$III)\,|S_Q\rangle = \frac{1}{\sqrt{2^l}}\left[|s_{Q1}\rangle|w_1\rangle + |s_{Q2}\rangle|w_2\rangle + \ldots + |s_{Ql}\rangle|w_l\rangle\right]. \qquad (4)$$

Equation (2) represents the situation in which there is no randomness, the compression is maximal. Equation (3) (that is a particular situation used to make easier the understanding), by its turn, represents a situation where there is some randomness and, hence, the compression is not maximal. In this example there are only two witnesses, thus the set $S_Q$ is partitioned in two parts. The extension for larger partitions is straightforward. At last, equation (4) represents the situation where there is maximal randomness and none compression. Observing (2)-(4) one can note a relation between the randomness and the entanglement between the quantum states of $|S_Q\rangle$ and $|W\rangle$. In equation (2) the knowledge of the value of $|w\rangle$ gives none information about the value of $|S_Q\rangle$ (any $S_{Qi}$ value is equally probable for $i = 1,\ldots,l$). In equation (3), the knowledge of $|w\rangle$ gives some information about the value of $|S_Q\rangle$ (for example, if $w = w_1$ then $S_Q$ belongs to the set $[S_{Q1},\ldots,S_{Qk}]$). Finally, in equation (4), the knowledge of the value of $|w\rangle$ gives complete information about the value of $|S_Q\rangle$ (for example, if $w = w_i$ then $S_Q = S_{Qi}$).